\documentclass[11pt,tightenlines,eqsecnum,floats,aps,nofootinbib,prd,showpacs,twocolumn]{revtex4}

\usepackage{amsmath,amssymb,amsfonts,amsthm,amscd}
\usepackage{graphicx}
\usepackage{enumerate}
\usepackage{colordvi}
\usepackage{units}
\usepackage{epsfig}
\usepackage{natbib}
\usepackage{enumerate}
\usepackage{colordvi}

\def\be{\begin{equation}}
\def\ee{\end{equation}}
\def\ba{\begin{eqnarray}}
\def\ea{\end{eqnarray}}

\begin{document}

\title{Bouncing Anisotropic Universes with Varying Constants}
\author{John D. Barrow$^1$}
\email{J.D.Barrow@damtp.cam.ac.uk}
\author{David Sloan$^1$}
\email{djs228@hermes.cam.ac.uk}
\affiliation{$^{1}$ DAMTP, Centre for Mathematical Sciences, Wilberforce Rd., Cambridge
University, Cambridge CB3 0WA, UK}

\begin{abstract}
We examine the evolution of a closed, homogeneous and anisotropic cosmology
subject to a variation of the fine structure 'constant', $\alpha $ within
the context of the theory introduced by Bekenstein, Sandvik, Barrow and
Magueijo, which generalises Maxwell's equations and general relativity. The
variation of $\alpha $ permits an effective ghost scalar field, whose
negative energy density becomes dominant at small length scales, leading to
a bouncing cosmology. A thermodynamically motivated coupling which describes
energy exchange between the effective ghost field and the radiation field
leads to an expanding, isotropizing sequence of bounces. In the absence of
entropy production we also find solutions with stable anisotropic
oscillations around a static universe.
\end{abstract}

\maketitle

\section{Introduction}

Spatially homogeneous cosmological models are a key area of study within
relativity. The introduction of anisotropies gives rise to models in which a
richer dynamical structure emerges, yet the cosmology remains simple enough
to provide analytic and simple numerical results. These models serve as a
test-bed for physical theories, and allow us analyse questions about why the
universe appears to be highly isotropic, whether inflation occurs for
generic or stable sets of initial data, the effects of anisotropy on
astronomical observables, and the behaviour of cosmological models on
approach to spacetime singularities \cite{Rendall:2005nf}, \cite%
{Andersson:2000cv}, \cite{Garfinkle:2012zh}.

The idea that the fine structure constant, $\alpha $, is a spacetime varying
scalar field was first investigated by Bekenstein \cite{Bekenstein:1982eu},
who created a natural generalisation of Maxwell's equations to accommodate a
varying electron charge. This idea was extended to include gravity and
provide a theory to explore cosmological consequences of varying $\alpha $
by Sandvik et al \cite{Sandvik:2001rv}. The resulting
Bekenstein-Sandvik-Barrow-Magueijo (BSBM) isotropic cosmological models were
found and used in conjunction with the astronomical data on varying $\alpha $
obtained from observations of high redshift quasar spectra \cite{Webb:2010hc}%
. More recently, the BSBM theory has been extended to included the case
where there is a coupling function (rather than simply a coupling constant)
between the charged matter fields and the scalar field driving changes in $%
\alpha $ \cite{Barrow:2011kr} and where that scalar field possesses a
self-interaction potential \cite{Barrow:2008ju}. These theories are the
analogues of the Jordan-Brans-Dicke theories for varying $G$ \cite%
{Brans:1961sx}.

In \cite{Barrow:2004ad} it was shown how theories of this type could produce
singularity-free homogeneous and isotropic cosmologies which displayed
stable oscillations around an Einstein static universe because the effect of
variations in the scalar field driving variations in $\alpha $ is to
introduce a negative 'ghost' density. Barrow and Tsagas \cite{Barrow:2009sj}
considered a broader context for these solutions and showed how the
inclusion of simple anisotropic expansion can modify the results because the
anisotropy can diverge just as quickly as a bounce-producing ghost scalar
field on a approach to the singularity. In this paper we will consider more
general closed anisotropic cosmologies with anisotropic 3-curvature in this
same context.

Matter bounces introduced by the presence of ghost fields are not a new
discovery (for a detailed examination see \cite{Gibbons:2003yj}). However,
in BSBM models the ghost field is an effective manifestation of underlying
physics, not a new matter source introduced by hand. In such models quantum
effects are ignored because of the prevailing attitude that ghost fields
should not be quantized (and are in fact ill-behaved when quantized, with
negative probability states). Furthermore, any coupling between a ghost
field and a non-ghost field would allow an infinite amount of energy to be
transferred from the ghost field.

We simply take the view that BSBM models can serve as test models for
bouncing cosmologies. The idea of a \textquotedblleft
phoenix\textquotedblright\ universe within relativity is almost as old as
big-bang models themselves, and goes back to Tolman \cite{Tolman} and Lema%
\^{\i}tre \cite{Lemaitre}. This classical picture of oscillating closed
universes with zero value of the cosmological constant, $\Lambda $, painted
by Tolman is well known. If there is no entropy production then cycles for
the time-evolution of the scale factor are periodic with the same amplitude
and total lifetime. If entropy increase is introduced in accord with the
second law of thermodynamics then the oscillating cycles become larger and
longer to the infinite future. The classical picture for isotropic universes
was competed by Barrow and D\c{a}browski \cite{BarrowDabrowski}, who showed
that if a positive cosmological constant is included then the sequence of
growing cycles will always come to an end, no matter how small the value of $%
\Lambda >0$. The ensuing behaviour will be to approach de Sitter expansion.
If the entropy increase from cycle to cycle is small then the asymptotic
state will be one in which the expansion is very close to a zero-curvature
state with comparable energy densities associated with matter and dark
energy (ie the cosmological constant). The dark energy will necessarily be
slightly dominant and the curvature will be positive -- not unlike the
situation in our observed universe. Barrow and D\c{a}browski  \cite%
{BarrowDabrowski} also considered the evolution of some simple bouncing
anisotropic universes of Kantowski-Sachs type, but not in a context that
included varying constants.

Many current quantum theories of gravity exhibit curvature singularity
avoidance, often in the form of a bounce (although they do not necessarily
avoid geodesic incompleteness). In Loop Quantum Cosmology, holonomy
corrections to the Friedmann equation give rise to a bounce at Planck scales
(see \cite{Ashtekar:2011ni} for a review). Horava-Lifschitz gravity also
introduces higher-order curvature corrections to Einstein's equations which
can cause the universe to bounce \cite{Brandenberger:2009yt} for some
parameter choices. In the latter case the dynamics of an anisotropic
solution have also been explored \cite{Misonoh:2011mn}.

The aim of this paper is to extend \cite{Barrow:2004ad} and \cite%
{BarrowDabrowski} to spatially homogeneous models which exhibit local
rotational symmetry (LRS) an so are effectively axisymmetric. LRS models
exhibit some of the features of full anisotropic model \cite{Calogero:2009mi}%
, yet the differential equations governing their dynamics can be solved with
relative ease using numerical solvers and exact methods. We will begin in
section \ref{PhysVar} by setting our the action principle underlying
variation of the fine structure 'constant', then in section \ref{CosSetup}
we set out the equations of motion for our system and define quantities of
physical interest, such as shear and Hubble expansion rates, in terms of
metric variables. Section \ref{Solutions} deals with two specific solutions
to the equations of motion: a static solution and ghost-induced inflation.
In particular, we will focus on the role played by anisotropies in both
these cases and examine perturbations about isotropic cases.

\section{Theories of Varying Alpha}

\label{PhysVar}

Varying \textquotedblleft constants \textquotedblright\ can be described by
extensions of the standard model of particle physics and/or general
relativity (GR) by the promotion of constants to space and time dependent
scalar fields. A well known and much studied example is that of
Jordan-Brans-Dicke theory in which GR is extended by generalising Newton's
constant $G$ to become a field variable \cite{Brans:1961sx}. These
self-consistent models for the variation of constants necessarily contain
conservation equations for the energy and momentum carried by the varying
scalar field and the gravitational field equations account for the scalar
field's effect on the spacetime geometry. This is in contrast to much of the
old literature on varying constants, other than $G,$ which merely 'write-in'
variations of constants into the equations that hold in the theory where the
constant does not vary. The existence of a self-consistent theory for the
variation of a constant also shows that much discussion about the meaning of
the variation of dimensional constants is not relevant because the solution
of the second-order conservation equation for the scalar field describing
the variation of a traditional constant always produces constants of
integration with the same dimensions as the varying constant and a
dimensionless combination is trivially available.

Physical models with extra dimensions often exhibit massless or light
degrees of freedom which can lead to the variation of such constants \cite%
{Damour:2003iz}, \cite{Uzan:2002vq} and there is the possibility for
observational bounds to be placed on any shift in the size of extra
dimensions over the age of the universe \cite{Kolb:1985sj},\cite%
{Barrow:2005sv}.

In this paper we shall follow the BSBM model in which the fine structure
constant, $\alpha $ is taken to be dynamical. Evidence of a dipolar spatial
variation has been recently claimed \cite{Webb:2010hc} and therefore it is
natural to extend this scenario to consider space and time variations but in
this paper we will only discuss time variations so that we can confine
attention to ordinary differential equations. Such variations are bounded by
terrestrial experiments to have small variation at present \cite%
{Marion:2002iw} \cite{Berengut:2010ht}. However evidence that the variation
is small currently does not rule out more significant changes in the past.
In particular, in BSBM theories $\alpha $ is not expected to vary during the
radiation era, to increase only logarithmically in time during the cold dark
matter dominated era, and then to become constant after the universal
expansion begins accelerating. Thus laboratory experiments today would not
be expected to find evidence for the variation of $\alpha $ found in
high-redshift quasar observations (that derive from epochs before the
universe began accelerating) even though it has been proved that any
cosmological variations in $\alpha $ will be seen in terrestrial experiments 
\cite{bshaw}.

The BSBM model describes the effect of varying the fine structure constant
by the introduction of a scalar dielectric field, $\psi $ with evolution of
the charge of an electron given by $e=e_{0}e^{\psi }$, in which $e_{0}$ is
the value of the electron charge at some fixed time, for example today.
Notice that $e_{0}$ is a fundamental constant and $e/e_{0}$ is
dimensionless. It has been shown \cite{Bekenstein:2009fq} that in spite of
modifying black hole solutions, the variability of $\alpha $ respects the
second law of thermodynamics. This will be important in section \ref%
{CosSetup}, as we will assume that all couplings between our fields obey the
second law.

The physical action is given by

\begin{equation}  \label{action}
S=\int \sqrt{-g} (\mathcal{L}_g +\mathcal{L}_m +L_\psi +e^{-2 \psi} \mathcal{%
L}_{em})
\end{equation}

where $\mathcal{L}_{g}=R/16\pi G$ is the usual Einstein-Hilbert Lagrangian, $%
\mathcal{L}_{\psi }=-\frac{\omega }{2}\partial _{u}\psi \partial ^{u}\psi $
governs the scalar dielectric field, $\psi $, $\mathcal{L}_{em}=-\frac{1}{4}%
f_{\mu \nu }f^{\mu \nu },$ and $\mathcal{L}_{m}$ is a matter Lagrangian
independent of $\psi $. Of particular importance to this paper is the
constant coupling parameter $\omega $ which we shall take to be negative, so
rendering $\psi $ an effective ghost scalar field. We do not consider the
generalised case where $\omega ^{\prime }(\psi )\neq 0$, see \cite%
{Barrow:2011kr}. From now on it will be convenient to simply consider the
effective field, rather than the underlying dielectric. This field is
massless, and it is clear from the action that its motion will be monotonic,
as will be that of the effective induced fine structure constant. In terms
of fluids, this field will appear to be stiff, with equation of state $%
p_{\psi }=\rho _{\psi }<0$.\ In a closed anisotropic cosmological model we
will allow energy exchange to occur between the $\psi $ field and an
equilibrium radiation field with equation of state $3p_{r}=\rho _{r}$, to
model an entropy increasing non-equilibrium process.

\section{Cosmological Expansion}

\label{CosSetup}

The physical system under consideration will consist of a homogeneous
anisotropic cosmology. For simplicity we will examine a system which is
locally rotationally symmetric, and use this to gain insight into the more
general case. For a concise review of these cosmological models, see \cite%
{Calogero:2009mi}. This model is general enough to contain the purely
general relativistic ingredient of anisotropic 3-curvature, which is missing
from the simple anisotropic models of Bianchi types I and V. It includes the
closed Bianchi type IX universe but only in the axisymmetric case where no
chaotic behaviour occurs. The LRS type IX metric is \emph{\ }

\begin{equation*}
ds^{2}=dt^{2}-h_{ij}\sigma ^{i}\sigma ^{j}
\end{equation*}%
where $\sigma ^{i}$ are the $SO(3)$ invariant 1-forms \cite{Taub:1951}

\begin{eqnarray*}
\sigma ^{1} &=&\cos \psi d\theta +\sin \psi \sin \theta d\phi \\
\sigma ^{2} &=&-\sin \psi d\theta +\cos \psi \sin \theta d\phi \\
\sigma ^{3} &=&d\psi +\cos \theta d\phi
\end{eqnarray*}%
and the LRS condition requires

\begin{equation*}
h_{ij}=diag\{a(t),b(t),b(t)\}
\end{equation*}

The metric contains two time-dependent scale factors (due to the LRS
condition), $a(t)$ and $b(t)$. The energy densities are denoted by $\rho _{r}
$ for radiation, $\rho _{\psi }$ for the scalar field, and $\rho _{\Lambda }$
for the cosmological constant, and the total density and pressure are $\rho $
and $p$, where

\begin{equation*}
\rho =\rho _{r}+\rho _{\psi }+\rho _{\Lambda }.
\end{equation*}%
All matter matter sources have isotropic pressures. The independent
variables are the principal 3-curvatures

\begin{center}
$R_{1}^{1}=\frac{a^{2}}{2b^{4}},$

$R_{2}^{2}=\frac{1}{b^{2}}-\frac{a^{2}}{2b^{4}};$
\end{center}

the mean Hubble expansion rate is defined by

\begin{center}
$H=\frac{1}{3}(\frac{\dot{a}}{a}+2\frac{\dot{b}}{b}),$
\end{center}

and the expansion shear scalar by

\begin{center}
$\sigma =\frac{1}{3}(\frac{\dot{b}}{b}-\frac{\dot{a}}{a})$
\end{center}

These variables are subject to a constraint equation (the generalized
Friedmann equation with $8\pi G=c=1$)

\begin{equation}
\rho =\frac{1}{b^{2}}-\frac{a^{2}}{4b^{4}}+2\frac{\dot{a}\dot{b}}{ab}+\frac{%
\dot{b}^{2}}{b^{2}}.  \label{Constraint}
\end{equation}

The remaining field equations are:

\begin{eqnarray}
\frac{\ddot{a}}{a} &=&-\frac{1}{2b^{4}}-2\frac{\dot{a}\dot{b}}{ab}+\frac{%
\rho -p}{2},  \label{HEOM} \\
\frac{\ddot{b}}{b} &=&\frac{a^{2}}{2b^{4}}-\frac{1}{b^{2}}-\frac{\dot{a}\dot{%
b}}{ab}-\frac{\dot{b}^{2}}{b^{2}}+\frac{\rho -p}{2},  \label{bEOM} \\
\dot{\sigma} &=&-3H\sigma +\frac{1}{3}(R_{1}-R_{2}),  \label{3} \\
\dot{H} &=&-H^{2}-2\sigma ^{2}-\frac{1}{6}(\rho +3p).
\end{eqnarray}

These reduce to the special case of the closed Friedmann universes when $a=b$%
. The shear does not evolve with $\sigma \propto (ab^{2})^{-1}$ as in
Bianchi type I models because of the 3-curvature anisotropy on the
right-hand side of eq. (\ref{3}). For ease of exposition, let us define the
variables $r=\frac{a}{b}$ and $H_{b}=\dot{b}/b.$The essential field
equations then simplify to

\begin{equation}
\frac{\ddot{r}}{r}=\frac{1-r^{2}}{b^{2}}-\frac{3\dot{r}\dot{b}}{rb},
\label{rEOM}
\end{equation}

\begin{eqnarray}
\sigma &=&-\frac{\dot{r}}{3r},  \label{HBEOM} \\
H &=&\frac{\dot{b}}{b}-3\sigma =H_{b}-3\sigma .  \label{H}
\end{eqnarray}

The continuity equation, which implies constraint conservation, is

\begin{equation}
\dot{\rho}+3H(\rho +p)=0.
\end{equation}

This governs the total energy density and pressure. Now, we introduce some
energy exchange between the fluids so that they obey

\begin{eqnarray}
\dot{\rho _{\psi }} &+&6H\rho _{\psi }=s,  \label{radEOM} \\
\dot{\rho _{r}} &+&4H\rho _{r}=-s,
\end{eqnarray}

where, $s$\ parametrises the flow of energy between the scalar field and
radiation. For our purposes, this will be taken to be of the form

\begin{center}
$s=-\rho _{\psi }\beta $
\end{center}

where

\begin{center}
$\beta =\beta _{0}+\beta _{H}H^{2}+\beta _{\sigma }\sigma ^{2}$\ 
\end{center}

can include a linear coupling $\beta _{0}$, plus possible bulk,\emph{\ }$%
\beta _{H}$, and shear, $\beta _{\sigma }$, viscous contributions. In
general, the $\beta $'s need not be constants.

The scalar dielectric field evolves according to

\begin{equation}
\ddot{\psi}+(3H+\beta )\dot{\psi}=0  \label{SFEOM}
\end{equation}%
and so

\begin{equation*}
\dot{\psi}\propto a^{-3}\exp [-\int \beta dt].
\end{equation*}

\section{Solutions}

\label{Solutions}

In what follows we consider the Bianchi IX case with no cosmological
constant ($\rho _{\Lambda }=0$). If the matter content is a perfect fluid
and obeys $\rho +3p>0$, then these universes expand from an initial
curvature singularity to a maximum size before collapsing back to a future
curvature singularity; the spacetime is past and future geodesically
incomplete. However, since the effect of varying the fine structure constant
is to produce a ghost scalar field with $\rho _{\psi }<0$ which will
dominate dynamics at small length scales, the energy condition is violated
and solutions to the BSBM model exist which have infinite past and future
temporal range.

In this section we will examine two particular solutions to the equations of
motion. The first is that of a static spacetime, and its behaviour under
perturbations. The second is that of an spacetime in which the coupling
between fields leads to inflationary behaviour.

\subsection{The Static Solution}

There exists a static solution of the form

\begin{equation}
\rho _{\psi }=-\frac{3}{4b^{2}} \:\:\: \rho _{r}=\frac{3}{2b^{2}},
\end{equation}%
for any given value of $b$. Note that in order to be static the solution
must be isotropic ($\sigma =0$), since from eq. (\ref{3}) the 3-curvatures
must match ($R_{1}^{1}=R_{2}^{2}$) for the shear to remain constant and eq. (%
\ref{H}) requires $\sigma =0$.

Now consider making a small perturbation about the isotropic solution by
introducing a small anisotropy: $r=1+\epsilon $ where $r=1$ represents
isotropy. From \ref{rEOM}, we find that

\begin{equation}
\ddot{\epsilon}=-3\epsilon H_{b}+\frac{2\epsilon +3\epsilon ^{2}+\epsilon
^{3}}{b^{2}}.  \label{epEOM}
\end{equation}

Without loss of generality, we take the unperturbed static solution to have $%
b=1$. If we introduce small parameters $\delta (t)$ and $\eta (t)$ such that 
$b=1+\eta $, and $\rho =3/2+\delta $, then to first order in our small
parameters:

\begin{equation}
\epsilon (t)=\epsilon _{0}\sin (\sqrt{2}t).  \label{epStatic}
\end{equation}

\begin{figure}[th]
\includegraphics[width=0.5\textwidth]{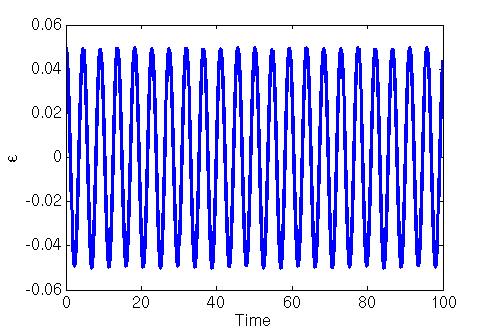}
\caption{$\protect\epsilon $ versus time with no coupling ($s=0$). Initial
values: $r=1.05,\protect\sigma =0,H=0.$}
\label{Epsoscill}
\end{figure}

In the case where there is no coupling between the fields ($s=0$), these
oscillations continue endlessly as shown in figure \ref{Epsoscill}. However,
once coupling is introduced ($s\neq 0$), the static case becomes unstable
because the balance between $\rho _{\psi }$ and $\rho _{r}$ is broken,
energy is transferred from the ghost field that supports stable oscillations
and eventually, after several oscillations, it settles into
radiation-dominated expansion, shown in figure \ref{Epsdecay}: 
\begin{figure}[th]
\includegraphics[width=0.5\textwidth]{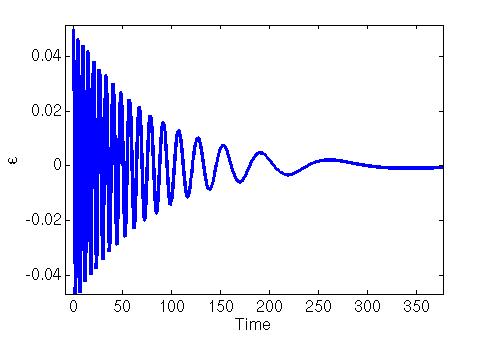}
\caption{$\protect\epsilon $ versus time with coupling turned on, $s\neq 0$,
showing the system isotropizing. Initial values: $r=1.05,\protect\sigma %
=0,H=0,\protect\beta _{0}=0.05$, $\protect\beta _{H}=\protect\beta _{\protect%
\sigma }=0.$}
\label{Epsdecay}
\end{figure}

From the constraint equation \ref{Constraint}, we find that to first order,
our small parameters are related by $2\delta =-\epsilon -3\eta $.
Furthermore, we can decompose $\delta $ into the radiation and scalar field
components, $\delta _{r}$ and $\delta _{\psi }$. In the absence of field
couplings, there is a relationship between these fields, due to their
coupled equations of motion \ref{SFEOM}, \ref{radEOM}:

\begin{equation}
\rho _{r}\propto \rho _{\psi }^{3/2}.  \label{MattRelations}
\end{equation}

In the static case under consideration, the constant of proportionality is $-%
\sqrt{6}$. Note that this relationship is broken by introducing a coupling
between the fields. For small $\delta $ we are therefore led to $\delta
_{r}=4\delta =-2\epsilon -6\eta $.

From \ref{bEOM}, the evolution of $\eta $ is given by:

\begin{equation}
\ddot{\eta}=\epsilon -\eta +\frac{\delta _{r}}{3}=\frac{\epsilon }{3}-\eta .
\label{etaEOM}
\end{equation}

Therefore there is a (more complicated) stable oscillatory behaviour for $%
\eta $ about the static solution and the evolution of $\epsilon $ has
already been determined by \ref{epStatic}. Thus we have an unusual behaviour
characterised by stable anisotropic oscillations around the isotropic
Einstein static universe. This generalises the simple isotropic oscillations
about the static universe that exist in Friedmann universes with a ghost
field found in \cite{Barrow:2004ad} and \cite{Barrow:2009sj}.

We can now determine further effects of allowing a coupling between the
fields. The second law of thermodynamics requires $s\geq 0$. The exact form
of $s$ - taking into account terms representing constant coupling, bulk and
shear viscosities, will of course affect the exact dynamics. However, it is
possible to make progress by assuming only that $s$ is non-negative. 

\begin{figure}[th]
\includegraphics[width=0.5\textwidth]{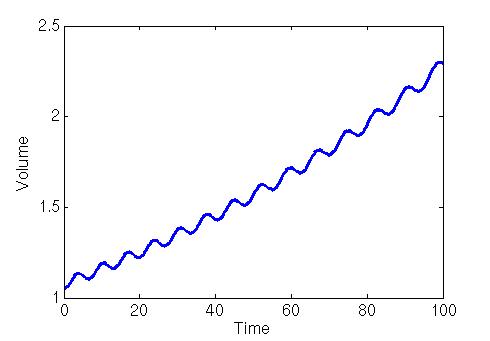}
\caption{Spatial volume versus time with coupling turned on ($s\neq 0$). The
system oscillates, but expands from one cycle to the next. Initial values: $%
r=1.05,\protect\sigma =0,H=0,\protect\beta _{0}=0.01$, $\protect\beta _{H}=%
\protect\beta _{\protect\sigma }=0.$}
\label{Volexpand}
\end{figure}

First consider the case of no coupling ($s=0$). The evolution of the
radiation field is determined by

\begin{equation}
\frac{\dot{\rho}_{r}}{\rho _{r}}=-\frac{4}{3}H=-\frac{4}{3}(\frac{\dot{%
\epsilon}}{3}+\dot{\eta}).  \label{RadCycle}
\end{equation}

Due to the cyclic behaviour of $\epsilon $ and $\eta $ (and hence of their
derivatives), $\delta _{r}$ will also cycle, returning to its initial value,
as any integral of the right-hand side of \ref{RadCycle} across a complete
cycle will be zero. However, with a positive coupling between the fields,
this relationship is broken, and a term which is always non-negative (and so
has a positive integral across cycles) must be added. Hence, across a cycle
in $\eta $ and $\epsilon ,$ the value of $\delta _{r}$ now increases and we
must adjust our equation of motion \ref{etaEOM} to include this term. We
write

\begin{equation}
\ddot{\eta}=\epsilon +\eta +\frac{\delta _{r}}{3}=\frac{\epsilon }{3}-\eta
+\Delta ,  \label{eta2EOM}
\end{equation}%
where $\Delta $ is a positive term representing the increase in $\delta _{r}$
due to field couplings created by introducing $s>0$. The variable $\eta $ no
longer cycles about zero, and the system is slowly pushed away from
stability, and enters a pseudo-cyclic phase in which a series of bounces
occur with increasing local  minima and maxima of the expansion volume, $%
ab^{2}$, as shown in figure \ref{Volexpand}. The behaviour of $\epsilon $ %
\ref{epEOM} is affected by this expansion (recall that $\epsilon $ is
already small). The expansion of $b(t)$ means that $H_{b}$ is no longer
small, and the equation gains a damping term. Similarly, the frequency of
oscillations,$\sqrt{2}/b,$ is reduced by this expansion in $b(t)$ and the
solution takes the approximate form of a damped harmonic oscillator. Note
that in the derivation of \ref{epEOM} only the smallness of $\epsilon $ was
used - therefore this damping behaviour is present in all expanding
solutions.

\subsection{Ghost-induced Inflation}

\begin{figure}[th]
\includegraphics[width=0.5\textwidth]{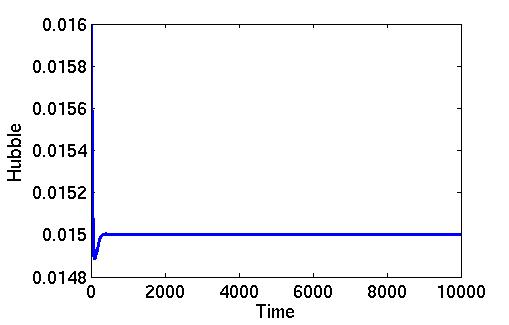}
\caption{Hubble expansion rate versus time for an inflating solution.
Initial values: $\protect\beta _{0}=0.03$, $\protect\beta _{H}=\protect\beta %
_{\protect\sigma }=0,$ $r=1,\protect\sigma =5\times
10^{-4},a=10^{4},H=1.6\times 10^{-2}.$}
\label{InflationFig}
\end{figure}

Spacetimes which exhibit inflation are of special interest to cosmologists
because inflation can solve a number of well known puzzles about the
universe's structure \cite{linde1986eternal}, and make a series of detailed
predictions that can be tested by observations of the microwave background
radiation \cite{Ade:2013xsa}. Typical inflationary models exhibit expansion
in which the Hubble parameter is (approximately) constant for a finite time
interval. In general relativistic cosmology this is usually achieved by
introducing a matter content that is (or is equivalent to) a scalar field
subject to a self-interaction potential whose contribution to the total
energy density is dominant during this expansion, with an equation of state
close to that of that produced by an exact cosmological constant with $%
p_{\Lambda }=-\rho _{\Lambda }$.

Inflation induced by ghost fields has been studied as an alternative to the
usual slow-roll models \cite{ArkaniHamed:2003uz}. Such models have
potentially observable consequences for the microwave background trispectrum 
\cite{Izumi:2010wm} \cite{Huang:2010ab}, but require that the translation
invariance of the scalar ghost field is broken. In the BSBM models under
consideration, however, translation invariance can be preserved, with the
field coupling responsible for creating the inflationary energy density.

Let us examine the case of a linear coupling-induced inflation with $s=\beta
_{0}\rho _{\psi },$ for constant $\beta _{0}>0$. When the volume is large
there exists an asymptotic solution of the form

\begin{equation}
\rho _{r}=\frac{9\beta _{0}^{2}}{4}, \:\:\: \rho _{\psi }=-\frac{3\beta
_{0}^{2}}{2},\:\:\: H=-\frac{\beta _{0}}{2},  \label{infcond}
\end{equation}%
in the isotropic case. This solution is stable, and is approached by
dynamical trajectories, as shown in figure \ref{InflationFig}. Under a small
perturbation $H=-\beta _{0}/2+h$ and $r=1+\epsilon $ we find that to first
order in the small parameters:

\begin{eqnarray}
\dot{h} &=&-\beta _{0}h  \label{infstab} \\
\ddot{\epsilon} &=&-3\dot{\epsilon}H-2\epsilon /b
\end{eqnarray}

In this solution, $b(t)$ is exponentially growing, so the final term in \ref%
{infstab} quickly becomes negligible. Therefore, although a shearing
expansion may occur, $\dot{\epsilon}$ quickly falls to zero, locking the
shear at a fixed value. This inflationary phenomenon is not unique to linear
couplings with $\beta _{H}=\beta _{\sigma }=0$, but is simplest to
demonstrate in this case. Likewise, there is no requirement for the matter
field to consist solely of radiation - introducing more matter fields with
couplings whose sign is determined to be in accordance with the second law
of thermodynamics yields a system which also exhibits inflation of this
type. For most couplings, however, the inflationary phase will end when the
dust field becomes dominant.

\begin{figure}[th]
\includegraphics[width=0.5\textwidth]{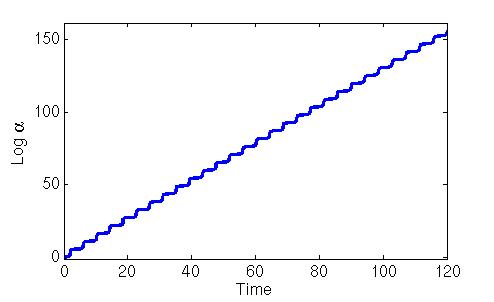} %
\includegraphics[width=0.5\textwidth]{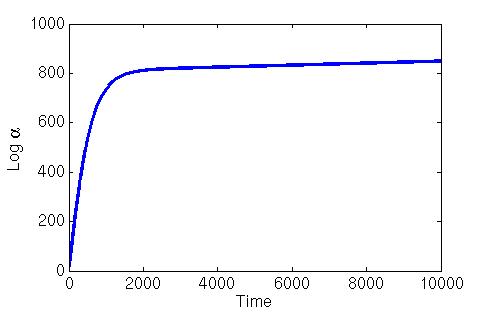}
\caption{Evolution of $\log (\protect\alpha )$ versus time shown over a
short timescale (top), and the same evolution over a long timescale
(bottom). Initial values: $\protect\beta _{0}=0.01$, $\protect\beta _{H}=%
\protect\beta _{\protect\sigma }=0,$ $r=1.01,\protect\sigma =0,a=1,H=0,%
\protect\rho _{\protect\psi }=-0.01.$}
\label{alphagraph}
\end{figure}

The evolution of the fine structure constant is shown in figure \ref%
{alphagraph}. Initially, the solution is like an ascending staircase with
rapid changes at each scale factor bounce, see for comparison \cite%
{Barrow:2004ad}. Monotonicity of $\alpha $ is ensured since the scalar field
cannot have positive energy density; since $\dot{\psi}=\sqrt{\frac{-2\rho
_{\psi }}{w}}$, we have $\dot{\psi}\geq 0$ for all time, and so $\psi $
cannot oscillate through maxima and minima. Across repeated bounces, $%
log(\alpha )$ will appear to increase in steps when $|\rho _{\psi }|$ is
small as the relative size of this energy density oscillates greatly within
a single cycle. However, as energy is transferred into the radiation field,
these steps will become less apparent, eventually approaching a constant
gradient once the ghost field reaches the condition for de Sitter inflation
to occur. Thus, even though the universe oscillates from cycle to cycle, the
fine structure 'constant' continues increasing from cycle to cycle and there
will typically only be a finite interval of cycles in which $\alpha $ takes
values that allow stable atoms to exist \cite{BarrowTipler},\cite%
{Lieb:1976zz}.

\section{Discussion}

In this paper, we examined the new effect of introducing anisotropies into
the BSBM framework for varying $\alpha $, although the conclusions have
broader applicability to anisotropic cosmologies containing ghost fields and
entropy-increasing energy exchanges between fields. In particular, we
studied the dynamics of locally rotationally symmetric Bianchi IX
cosmologies. It was shown that under certain conditions the bouncing
behaviour observed in isotropic models persists, with the fine structure
constant\textquotedblright\ changing in an almost step-like increasing
manner between cycles as time increases. It is apparent from \ref{HEOM} that
on short scales there is a tension between the shear terms and ghost field,
as both scale as the inverse square of the volume. When the anisotropy is
small, the contribution from the ghost field dominates. This leads to a
bouncing model, reproducing closely the results seen in \cite{Barrow:2004ad}%
. Furthermore, there exists a static solution, perturbations about which
lead to a sequence of anisotropic bouncing phases. When there is a coupling
between the matter fields, the second law of thermodynamics ensures that
this process isotropizes the system by energy exchange.

The resulting dynamics lead to a pseudo-cyclic universe in which the fine
structure constant monotonically increases across bounces, and for small
values of the associated dielectric scalar field, this increase is dominated
by dynamics near the bounce point. The minimum and maximum volumes of the
universe also increase across cycles, with the total energy density
decreasing. Eventually, the model reaches a point at which the coupling
between fields fixes the energy density to be constant in time, and the
universe undergoes a de Sitter phase in which it inflates. Since this model
is limited to include only the dielectric field and radiation, there is no
transition to the dust-dominated era that one would expect at the end of
this phase, and so inflation is endless within the model. It is possible to
find solutions with dust in which the system again reaches a point of steady
inflation. However within the space of couplings with $s>0$ which obey the
second law, these solutions are a set of measure zero.

The evolution of $\alpha $ throughout the history of BSBM universes displays
interesting traits. At late times, in a large universe it will appear that $%
\alpha $ has settled to a constant value. In doing so, throughout a series
of oscillations the universe will have isotropised greatly, with $\alpha $
stepping up between cycles. As the dynamics are invariant under changing the
initial value of $\alpha $ there is no obvious mechanism to determine the
constant to which it will approach.  

\section*{Acknowledgements}

DS acknowledges support from a Templeton Foundation grant.  

\bibliographystyle{apsrev}

\begin{thebibliography}{99}
\expandafter\ifx\csname natexlab\endcsname\relax\def\natexlab#1{#1}\fi
\expandafter\ifx\csname bibnamefont\endcsname\relax
  \def\bibnamefont#1{#1}\fi
\expandafter\ifx\csname bibfnamefont\endcsname\relax
  \def\bibfnamefont#1{#1}\fi
\expandafter\ifx\csname citenamefont\endcsname\relax
  \def\citenamefont#1{#1}\fi
\expandafter\ifx\csname url\endcsname\relax
  \def\url#1{\texttt{#1}}\fi
\expandafter\ifx\csname urlprefix\endcsname\relax\def\urlprefix{URL }\fi
\providecommand{\bibinfo}[2]{#2}
\providecommand{\eprint}[2][]{\url{#2}}


\bibitem[{\citenamefont{Rendall}(2005)}]{Rendall:2005nf}
\bibinfo{author}{\bibfnamefont{A.~D.} \bibnamefont{Rendall}},
  \emph{\bibinfo{title}{{The Nature of spacetime singularities}}}
  (\bibinfo{publisher}{World Scientific}, \bibinfo{address}{Washington, D.C.},
  \bibinfo{year}{2005}).

\bibitem[{\citenamefont{Andersson and Rendall}(2001)}]{Andersson:2000cv}
\bibinfo{author}{\bibfnamefont{L.}~\bibnamefont{Andersson}} \bibnamefont{and}
  \bibinfo{author}{\bibfnamefont{A.~D.} \bibnamefont{Rendall}},
  \bibinfo{journal}{Commun. Math. Phys.} \textbf{\bibinfo{volume}{218}},
  \bibinfo{pages}{479} (\bibinfo{year}{2001}).

\bibitem[{\citenamefont{Garfinkle}(2012)}]{Garfinkle:2012zh}
\bibinfo{author}{\bibfnamefont{D.}~\bibnamefont{Garfinkle}},
  \bibinfo{journal}{Class. Quant. Grav.} \textbf{\bibinfo{volume}{29}},
  \bibinfo{pages}{244003} (\bibinfo{year}{2012}).

\bibitem[{\citenamefont{Bekenstein}(1982)}]{Bekenstein:1982eu}
\bibinfo{author}{\bibfnamefont{J.}~\bibnamefont{Bekenstein}},
  \bibinfo{journal}{Phys. Rev.} \textbf{\bibinfo{volume}{D25}},
  \bibinfo{pages}{1527} (\bibinfo{year}{1982}).

\bibitem[{\citenamefont{Sandvik et~al.}(2002)\citenamefont{Sandvik, Barrow, and
  Magueijo}}]{Sandvik:2001rv}
\bibinfo{author}{\bibfnamefont{H.~B.} \bibnamefont{Sandvik}},
  \bibinfo{author}{\bibfnamefont{J.~D.} \bibnamefont{Barrow}},
  \bibnamefont{and} \bibinfo{author}{\bibfnamefont{J.}~\bibnamefont{Magueijo}},
  \bibinfo{journal}{Phys. Rev. Lett.} \textbf{\bibinfo{volume}{88}},
  \bibinfo{pages}{031302} (\bibinfo{year}{2002}).

\bibitem[{\citenamefont{Webb et~al.}(2011)\citenamefont{Webb, King, Murphy,
  Flambaum, Carswell et~al.}}]{Webb:2010hc}
\bibinfo{author}{\bibfnamefont{J.}~\bibnamefont{Webb}},
  \bibinfo{author}{\bibfnamefont{J.}~\bibnamefont{King}},
  \bibinfo{author}{\bibfnamefont{M.}~\bibnamefont{Murphy}},
  \bibinfo{author}{\bibfnamefont{V.}~\bibnamefont{Flambaum}},
  \bibinfo{author}{\bibfnamefont{R.}~\bibnamefont{Carswell}},
  \bibnamefont{et~al.}, \bibinfo{journal}{Phys. Rev. Lett.}
  \textbf{\bibinfo{volume}{107}}, \bibinfo{pages}{191101}
  (\bibinfo{year}{2011}).

\bibitem[{\citenamefont{Barrow and Lip}(2012)}]{Barrow:2011kr}
\bibinfo{author}{\bibfnamefont{J.~D.} \bibnamefont{Barrow}} \bibnamefont{and}
  \bibinfo{author}{\bibfnamefont{S.~Z.} \bibnamefont{Lip}},
  \bibinfo{journal}{Phys. Rev.} \textbf{\bibinfo{volume}{D85}},
  \bibinfo{pages}{023514} (\bibinfo{year}{2012}).

\bibitem[{\citenamefont{Barrow and Li}(2008)}]{Barrow:2008ju}
\bibinfo{author}{\bibfnamefont{J.~D.} \bibnamefont{Barrow}} \bibnamefont{and}
  \bibinfo{author}{\bibfnamefont{B.}~\bibnamefont{Li}}, \bibinfo{journal}{Phys.
  Rev.} \textbf{\bibinfo{volume}{D78}}, \bibinfo{pages}{083536}
  (\bibinfo{year}{2008}).

\bibitem[{\citenamefont{Brans and Dicke}(1961)}]{Brans:1961sx}
\bibinfo{author}{\bibfnamefont{C.}~\bibnamefont{Brans}} \bibnamefont{and}
  \bibinfo{author}{\bibfnamefont{R.}~\bibnamefont{Dicke}},
  \bibinfo{journal}{Phys. Rev.} \textbf{\bibinfo{volume}{124}},
  \bibinfo{pages}{925} (\bibinfo{year}{1961}).

\bibitem[{\citenamefont{Barrow et~al.}(2004)\citenamefont{Barrow, Kimberly, and
  Magueijo}}]{Barrow:2004ad}
\bibinfo{author}{\bibfnamefont{J.~D.} \bibnamefont{Barrow}},
  \bibinfo{author}{\bibfnamefont{D.}~\bibnamefont{Kimberly}}, \bibnamefont{and}
  \bibinfo{author}{\bibfnamefont{J.}~\bibnamefont{Magueijo}},
  \bibinfo{journal}{Class. Quant. Grav.} \textbf{\bibinfo{volume}{21}},
  \bibinfo{pages}{4289} (\bibinfo{year}{2004}).

\bibitem[{\citenamefont{Barrow and Tsagas}(2009)}]{Barrow:2009sj}
\bibinfo{author}{\bibfnamefont{J.~D.} \bibnamefont{Barrow}} \bibnamefont{and}
  \bibinfo{author}{\bibfnamefont{C.~G.} \bibnamefont{Tsagas}},
  \bibinfo{journal}{Class. Quant. Grav.} \textbf{\bibinfo{volume}{26}},
  \bibinfo{pages}{195003} (\bibinfo{year}{2009}).

\bibitem[{\citenamefont{Gibbons}(2003)}]{Gibbons:2003yj}
\bibinfo{author}{\bibfnamefont{G.}~\bibnamefont{Gibbons}}
  (\bibinfo{year}{2003}), \eprint{hep-th/0302199}.

\bibitem[{\citenamefont{{Tolman}}(1934)}]{Tolman}
\bibinfo{author}{\bibfnamefont{R.~C.} \bibnamefont{{Tolman}}},
  \emph{\bibinfo{title}{{Relativity, Thermodynamics, and Cosmology}}}
  (\bibinfo{publisher}{Clarendon Press}, \bibinfo{address}{Oxford},
  \bibinfo{year}{1934}).

\bibitem[{\citenamefont{Lema\^{i}tre}(1933)}]{Lemaitre}
\bibinfo{author}{\bibfnamefont{G.}~\bibnamefont{Lema\^{i}tre}},
  \bibinfo{journal}{Ann. de la Soc. Scientifique de Bruxelles}
  \textbf{\bibinfo{volume}{A53}}, \bibinfo{pages}{51} (\bibinfo{year}{1933}).

\bibitem[{\citenamefont{Barrow and D\c{a}browski}(1995)}]{BarrowDabrowski}
\bibinfo{author}{\bibfnamefont{J.~D.} \bibnamefont{Barrow}} \bibnamefont{and}
  \bibinfo{author}{\bibfnamefont{M.~P.} \bibnamefont{D\c{a}browski}},
  \bibinfo{journal}{Mon. Not. R. Astron. Soc.} \textbf{\bibinfo{volume}{275}},
  \bibinfo{pages}{850} (\bibinfo{year}{1995}).

\bibitem[{\citenamefont{Ashtekar and Singh}(2011)}]{Ashtekar:2011ni}
\bibinfo{author}{\bibfnamefont{A.}~\bibnamefont{Ashtekar}} \bibnamefont{and}
  \bibinfo{author}{\bibfnamefont{P.}~\bibnamefont{Singh}},
  \bibinfo{journal}{Class. Quant. Grav.} \textbf{\bibinfo{volume}{28}},
  \bibinfo{pages}{213001} (\bibinfo{year}{2011}).

\bibitem[{\citenamefont{Brandenberger}(2009)}]{Brandenberger:2009yt}
\bibinfo{author}{\bibfnamefont{R.}~\bibnamefont{Brandenberger}},
  \bibinfo{journal}{Phys. Rev.} \textbf{\bibinfo{volume}{D80}},
  \bibinfo{pages}{043516} (\bibinfo{year}{2009}).

\bibitem[{\citenamefont{Misonoh et~al.}(2011)\citenamefont{Misonoh, Maeda, and
  Kobayashi}}]{Misonoh:2011mn}
\bibinfo{author}{\bibfnamefont{Y.}~\bibnamefont{Misonoh}},
  \bibinfo{author}{\bibfnamefont{K.-i.} \bibnamefont{Maeda}}, \bibnamefont{and}
  \bibinfo{author}{\bibfnamefont{T.}~\bibnamefont{Kobayashi}},
  \bibinfo{journal}{Phys. Rev.} \textbf{\bibinfo{volume}{D84}},
  \bibinfo{pages}{064030} (\bibinfo{year}{2011}).

\bibitem[{\citenamefont{Calogero and Heinzle}(2011)}]{Calogero:2009mi}
\bibinfo{author}{\bibfnamefont{S.}~\bibnamefont{Calogero}} \bibnamefont{and}
  \bibinfo{author}{\bibfnamefont{J.~M.} \bibnamefont{Heinzle}},
  \bibinfo{journal}{Physica} \textbf{\bibinfo{volume}{D240}},
  \bibinfo{pages}{636} (\bibinfo{year}{2011}).

\bibitem[{\citenamefont{Damour}(2003)}]{Damour:2003iz}
\bibinfo{author}{\bibfnamefont{T.}~\bibnamefont{Damour}}, in
  \emph{\bibinfo{booktitle}{Neutrino telescopes. Proceedings}}, edited by
  \bibinfo{editor}{\bibfnamefont{M.}~\bibnamefont{Baldo-Ceolin}}
  (\bibinfo{publisher}{Instituto Veneto Di Scienze}, \bibinfo{address}{Venice},
  \bibinfo{year}{2003}), pp. \bibinfo{pages}{595--609}.

\bibitem[{\citenamefont{Uzan}(2003)}]{Uzan:2002vq}
\bibinfo{author}{\bibfnamefont{J.-P.} \bibnamefont{Uzan}},
  \bibinfo{journal}{Rev. Mod. Phys.} \textbf{\bibinfo{volume}{75}},
  \bibinfo{pages}{403} (\bibinfo{year}{2003}).

\bibitem[{\citenamefont{Kolb et~al.}(1986)\citenamefont{Kolb, Perry, and
  Walker}}]{Kolb:1985sj}
\bibinfo{author}{\bibfnamefont{E.~W.} \bibnamefont{Kolb}},
  \bibinfo{author}{\bibfnamefont{M.~J.} \bibnamefont{Perry}}, \bibnamefont{and}
  \bibinfo{author}{\bibfnamefont{T.}~\bibnamefont{Walker}},
  \bibinfo{journal}{Phys. Rev.} \textbf{\bibinfo{volume}{D33}},
  \bibinfo{pages}{869} (\bibinfo{year}{1986}).

\bibitem[{\citenamefont{Barrow}(2005)}]{Barrow:2005sv}
\bibinfo{author}{\bibfnamefont{J.~D.} \bibnamefont{Barrow}},
  \bibinfo{journal}{Phys. Rev.} \textbf{\bibinfo{volume}{D71}},
  \bibinfo{pages}{083520} (\bibinfo{year}{2005}).

\bibitem[{\citenamefont{Marion et~al.}(2003)\citenamefont{Marion, Pereira
  Dos~Santos, Abgrall, Zhang, Sortais et~al.}}]{Marion:2002iw}
\bibinfo{author}{\bibfnamefont{H.}~\bibnamefont{Marion}},
  \bibinfo{author}{\bibfnamefont{F.}~\bibnamefont{Pereira Dos~Santos}},
  \bibinfo{author}{\bibfnamefont{M.}~\bibnamefont{Abgrall}},
  \bibinfo{author}{\bibfnamefont{S.}~\bibnamefont{Zhang}},
  \bibinfo{author}{\bibfnamefont{Y.}~\bibnamefont{Sortais}},
  \bibnamefont{et~al.}, \bibinfo{journal}{Phys. Rev. Lett.}
  \textbf{\bibinfo{volume}{90}}, \bibinfo{pages}{150801}
  (\bibinfo{year}{2003}).

\bibitem[{\citenamefont{Berengut and Flambaum}(2012)}]{Berengut:2010ht}
\bibinfo{author}{\bibfnamefont{J.}~\bibnamefont{Berengut}} \bibnamefont{and}
  \bibinfo{author}{\bibfnamefont{V.}~\bibnamefont{Flambaum}},
  \bibinfo{journal}{Europhys. Lett.} \textbf{\bibinfo{volume}{97}},
  \bibinfo{pages}{20006} (\bibinfo{year}{2012}).

\bibitem[{\citenamefont{Shaw and Barrow}(2006)}]{bshaw}
\bibinfo{author}{\bibfnamefont{D.}~\bibnamefont{Shaw}} \bibnamefont{and}
  \bibinfo{author}{\bibfnamefont{J.}~\bibnamefont{Barrow}},
  \bibinfo{journal}{Phys. Rev.} \textbf{\bibinfo{volume}{D73}},
  \bibinfo{pages}{123505} (\bibinfo{year}{2006}).
\bibinfo{author}{\bibfnamefont{D.}~\bibnamefont{Shaw}} \bibnamefont{and}
  \bibinfo{author}{\bibfnamefont{J.}~\bibnamefont{Barrow}},
  \bibinfo{journal}{Phys. Rev.} \textbf{\bibinfo{volume}{D73}},
  \bibinfo{pages}{123506} (\bibinfo{year}{2006}).


\bibitem[{\citenamefont{Bekenstein and Schiffer}(2009)}]{Bekenstein:2009fq}
\bibinfo{author}{\bibfnamefont{J.~D.} \bibnamefont{Bekenstein}}
  \bibnamefont{and} \bibinfo{author}{\bibfnamefont{M.}~\bibnamefont{Schiffer}},
  \bibinfo{journal}{Phys. Rev.} \textbf{\bibinfo{volume}{D80}},
  \bibinfo{pages}{123508} (\bibinfo{year}{2009}).

\bibitem[{\citenamefont{Taub}(1951)}]{Taub:1951}
\bibinfo{author}{\bibfnamefont{A.~H.} \bibnamefont{Taub}},
  \bibinfo{journal}{Ann. Math.} \textbf{\bibinfo{volume}{53}},
  \bibinfo{pages}{472} (\bibinfo{year}{1951}).

\bibitem[{\citenamefont{Linde}(1986)}]{linde1986eternal}
\bibinfo{author}{\bibfnamefont{A.}~\bibnamefont{Linde}}, \bibinfo{journal}{Mod.
  Phys. Lett.} \textbf{\bibinfo{volume}{A1}}, \bibinfo{pages}{81}
  (\bibinfo{year}{1986}).

\bibitem[{\citenamefont{Ade et~al.}(2013)}]{Ade:2013xsa}
\bibinfo{author}{\bibfnamefont{P.~A.~R.} \bibnamefont{Ade}}
  \bibnamefont{et~al.} (\bibinfo{collaboration}{Planck Collaboration})
  (\bibinfo{year}{2013}), \eprint{arXiv:1303.5062 [astro-ph.CO]}.

\bibitem[{\citenamefont{Arkani-Hamed et~al.}(2004)\citenamefont{Arkani-Hamed,
  Creminelli, Mukohyama, and Zaldarriaga}}]{ArkaniHamed:2003uz}
\bibinfo{author}{\bibfnamefont{N.}~\bibnamefont{Arkani-Hamed}},
  \bibinfo{author}{\bibfnamefont{P.}~\bibnamefont{Creminelli}},
  \bibinfo{author}{\bibfnamefont{S.}~\bibnamefont{Mukohyama}},
  \bibnamefont{and}
  \bibinfo{author}{\bibfnamefont{M.}~\bibnamefont{Zaldarriaga}},
  \bibinfo{journal}{JCAP} \textbf{\bibinfo{volume}{0404}}, \bibinfo{pages}{001}
  (\bibinfo{year}{2004}).

\bibitem[{\citenamefont{Izumi and Mukohyama}(2010)}]{Izumi:2010wm}
\bibinfo{author}{\bibfnamefont{K.}~\bibnamefont{Izumi}} \bibnamefont{and}
  \bibinfo{author}{\bibfnamefont{S.}~\bibnamefont{Mukohyama}},
  \bibinfo{journal}{JCAP} \textbf{\bibinfo{volume}{1006}}, \bibinfo{pages}{016}
  (\bibinfo{year}{2010}).

\bibitem[{\citenamefont{Huang}(2010)}]{Huang:2010ab}
\bibinfo{author}{\bibfnamefont{Q.-G.} \bibnamefont{Huang}},
  \bibinfo{journal}{JCAP} \textbf{\bibinfo{volume}{1007}}, \bibinfo{pages}{025}
  (\bibinfo{year}{2010}).

\bibitem[{\citenamefont{Barrow and Tipler}(1986)}]{BarrowTipler}
\bibinfo{author}{\bibfnamefont{J.~D.} \bibnamefont{Barrow}} \bibnamefont{and}
  \bibinfo{author}{\bibfnamefont{F.~J.} \bibnamefont{Tipler}},
  \emph{\bibinfo{title}{The Anthropic Cosmological Principle}}
  (\bibinfo{publisher}{Oxford University Press}, \bibinfo{address}{Oxford},
  \bibinfo{year}{1986}).

\bibitem[{\citenamefont{Lieb}(1976)}]{Lieb:1976zz}
\bibinfo{author}{\bibfnamefont{E.~H.} \bibnamefont{Lieb}},
  \bibinfo{journal}{Rev. Mod. Phys.} \textbf{\bibinfo{volume}{48}},
  \bibinfo{pages}{553} (\bibinfo{year}{1976}).

\end{thebibliography}

\end{document}